\documentclass{aa}
\usepackage{graphicx}
\usepackage{supertabular}
\usepackage{txfonts}
\usepackage{natbib}
\usepackage{lscape}
\begin{document}
\title{Infrared photometry and evolution of mass-losing AGB stars.}
\subtitle{III. Mass loss rates of MS and S stars}

\author
    {
      R. Guandalini\inst{1}
     }

\offprints{R. Guandalini, guandalini@fisica.unipg.it}

\institute
    {Department of Physics, University of Perugia and INFN, Section of Perugia, Via A. Pascoli,
      06123 Perugia, Italy\\
      \email{guandalini@fisica.unipg.it}
    }

\date{Received / Accepted }

% \abstract{}{}{}{}{}
% 5 {} token are mandatory

\abstract
% context heading
{The asymptotic giant branch (AGB) phase marks the end of the
evolution for low- and intermediate-mass stars, which are
fundamental contributors to the mass return to the interstellar
medium and to the chemical evolution of galaxies. The detailed
understanding of mass loss processes is hampered by the poor
knowledge of the luminosities and distances of AGB stars.}
% aims heading
{In a series of papers we are trying to establish criteria
permitting a more quantitative determination of luminosities for
the various types of AGB stars, using the infrared (IR) fluxes as a
basis. An updated compilation of the mass loss rates
is also required, as it is crucial in our studies of the evolutionary
properties of these stars. In this paper we concentrate our analysis on the
study of the mass loss rates for a sample of galactic S stars.}
% methods heading
{We reanalyze the properties of the stellar winds for a sample of
galactic MS, S, SC stars with reliable estimates of the distance
on the basis of criteria previously determined. We then compare
the resulting mass loss rates with those previously obtained
for a sample of C-rich AGB stars.}
%results heading
{Stellar winds in S stars are on average less efficient than those
of C-rich AGB stars of the same luminosity. Near-to-mid
infrared colors appear to be crucial in our analysis. They show a
good correlation with mass loss rates in particular for the Mira stars.
We suggest that the relations between the rates of the stellar winds
and both the near-to-mid infrared colors and the periods of
variability improve the understanding of the late evolutionary stages of
low mass stars and \textbf{could be the} origin of the relation between the
rates of the stellar winds and the bolometric magnitudes.}{}


\titlerunning{The Evolutionary Status of Mass-Losing AGB Stars. III.}
\authorrunning{R. Guandalini}

\maketitle

%________________________________________________________________

\section{Introduction \label{sect1}}

Stars of low and intermediate mass (between $\sim$~0.8 and
8.0\,M$_{\odot}$) terminate their "active" evolution with the
asymptotic giant branch (hereafter AGB) phase through the
alternate burning of two nuclear shells of H and He. Detailed
reviews on the main nuclear and evolutionary properties of AGB
stars are presented in \citet{busso99,herwig} and references
therein.

On the AGB, S-type stars are identified spectroscopically through
the detection of enhanced s-process abundances. This is found in
two classes of objects called "extrinsic" and "intrinsic" S stars.
A star showing s-process elements after phenomena of mass-transfer
in a binary system is called extrinsic. An intrinsic AGB star on
the other hand brings s-elements to the surface through repeated third
dredge-up episodes along the AGB phase. Sometimes this is revealed
only by the presence of \textbf{Technetium}, while other signatures of s-elements
remain hard to detect \citep{utt07}.

Important gaps in the knowledge of crucial physical parameters
still undermine our understanding of the whole evolutionary
sequence along the AGB: this is particularly so for luminosities and
mass loss rates. AGB stars lose substantial amounts of matter, as
their winds are the main contributors for the replenishment
of the interstellar medium \citep{sedlmayr94}. Various attempts
were made to describe the mass loss mechanism
\citep[see e.g.][]{salpeter,knappmorris,wachter02}.
Unfortunately, quantitative knowledge of stellar winds is still
poor, forcing the adoption of parametric treatments where
observations (at infrared or radio wavelengths) and distance
estimates play a crucial role.

Similar problems also affect the other crucial parameter: the
bolometric luminosity. The difficult estimates of distance have a
large influence also in this case. Moreover, the luminosities
derived from full stellar evolutionary models are affected by many
uncertainties in the choice of their parameters \citep[see][and
references therein]{straniero}.

%\begin{landscape}
\begin{table*}[t!]
\begin{minipage}[t]{\textwidth}
\caption{Relevant data for our sample of S-MS stars.}
% title of Table
\label{table:1}      % is used to refer this table in the text
\centering                          % used for centering table
\renewcommand{\footnoterule}{}  % to avoid a line before footnotes
\begin{tabular}{l c c c c r r c c c}        % centered columns (4 columns)
\hline\hline
Source   & Spectral Type & Var. Type &  Distance\footnote{References
for the distances are: \emph{A}) the revised Hipparcos catalogue
\citep{vleu2007}; \emph{B}) the period-luminosity methods for the
O-rich stars as used in Paper II.}   & Bol. Magnitudes\footnote{Bolometric Magnitudes are taken from Paper II.} &   Mass Loss   &  v$_{e}$ &   Ref.\footnote{References for the mass loss
rates (updated with our choice for distances) and for the outflow velocities are quoted as: R
stands for \citet{ramstedt09}; W is for \citet{winters03}; L for
\citet{loup}.} &  I. $-$ E. & dust/gas \\
Name   & & (GCVS) &  (kpc)   & (Paper II) &   (M$_{\odot}$/yr)  & km/s &   &   & ratio \\
\hline\hline
\object{S Cas}  & S3,4e$-$S5,8e & Mira &   0.85$^B$    & $-$5.71 &   1.02E$-$5 & 20.5   &   R & I & 2.16E$-$4  \\
\object{W Aql}  & S3,9e$-$S6,9e & Mira &   0.34$^B$    & $-$5.44  &   4.24E$-$6    &   17.2 & R   & I & 5.66E$-$4 \\
\object{R Cyg}  & S2.5,9e$-$S6,9e(Tc) & Mira &   0.55$^B$    & $-$5.42  &   9.04E$-$7    &   9.0 & R   & $-$ & 1.18E$-$3  \\
\object{chi Cyg}    & S6,2e$-$S10,4e/MSe & Mira &   0.18$^A$    & $-$5.39  &   8.69E$-$7    & 8.5   &   R   & I & 1.15E$-$4  \\
\object{T Cam}  & S4,7e$-$S8.5,8e & Mira &   0.50$^B$ & $-$5.22 & 8.91E$-$8    &   3.8   &   R & I & 6.73E$-$4 \\
\object{R Lyn} & S2.5,5e$-$S6,8e: & Mira & 0.95$^B$ & $-$5.19 & 3.90E$-$7    &   7.5   & R   & I & 3.70E$-$4 \\
\object{R Gem}  & S2,9e$-$S8,9e(Tc) & Mira &   0.66$^B$    & $-$5.23 &   3.91E$-$7    &   4.5 & R   & I & 2.67E$-$4  \\
\object{GI Lup} & S7,8e & Mira &   0.80$^B$    & $-$ &   7.20E$-$7    &   10.0 &   R   & I & 7.56E$-$4 \\
\object{WY Cas} & S6,5pe & Mira &   0.97$^B$    & $-$5.50 &   2.26E$-$6    &   13.5 &   R   & I & 1.56E$-$3 \\
\object{pi1 Gru}    & S5,7e & SRB &   0.16$^A$    & $-$5.75 &   2.57E$-$6 &   14.5 &   W   & I & 3.71E$-$4 \\
\object{ST Her} & M6$-$7IIIaS & SRB & 0.30$^A$ & $-$5.64 &   1.30E$-$7 &   8.5  &   R   & I & 4.62E$-$3 \\
\object{T Cet}  & M5$-$6SIIe & SRC &   0.27$^A$ & $-$5.63 & 4.93E$-$8    &   5.5 &   R & I & 2.08E$-$3 \\
\object{Y Lyn}  & M6SIb$-$II & SRC &   0.25$^A$    & $-$5.33 &   2.17E$-$7    &   7.5 & R & I & 7.83E$-$4  \\
\object{R And}  & S3,5e$-$S8,8e/M7e & Mira &   0.41$^B$    & $-$5.19 &   1.09E$-$6    & 8.3   &   R   & I & 3.67E$-$4 \\
\object{W And}  & S6,1e$-$S9,2e/M4$-$M1 & Mira &   0.38$^B$    & $-$5.27 &   2.79E$-$7    &   6.0 &   R   & I & 1.64E$-$3 \\
\object{RT Sco} & S7,2/M6e$-$M7e & Mira &   0.45$^B$ & $-$5.44 & 1.01E$-$6    &   11.0   & R   & I & 9.84E$-$4 \\
\object{RS Cnc} & M6eIb$-$II/S & SRC &   0.14$^A$    & $-$5.21  & 2.80E$-$7 &   6.8 &   L   & I & 6.78E$-$4 \\
\object{AA Cam} & M5/S & LB & 0.78$^A$    & $-$ & 3.83E$-$8    &   3.4   &   R & I & 2.10E$-$2 \\
\hline
\object{S Lyr}  & SCe & Mira &   2.27$^B$    & $-$5.50 & 5.44E$-$6    & 13.0  &   R   & I & 4.41E$-$4 \\
\object{TT Cen} & CSe & Mira &   1.39$^B$    & $-$5.48 &   5.17E$-$6    &   20.0 & R   & $-$ & 2.01E$-$3 \\
\object{ST Sgr} & C4,3e$-$S9,5e & Mira &   0.76$^B$ & $-$5.24 &   3.44E$-$7    &   6.0   & R   & I & 1.51E$-$3 \\
\object{UY Cen} & SC & SR &   0.69$^A$ & $-$6.05  & 1.70E$-$7    &   12.0 &   R & I & 2.35E$-$3 \\
\hline
\object{VX Aql} & C9,1p/M0ep & Mira & 1.99$^B$ & $-$5.87  & 3.51E$-$7    &   7.0 & R & I & 4.40E$-$3 \\
\hline \hline
\end{tabular}
\end{minipage}
\end{table*}
%\end{landscape}

We are performing an analysis of ground-based and
space-borne IR observations of AGB stars trying to reduce the
uncertainties still present in the determination of those two
parameters and in their relations with the others, thus obtaining
improved evolutionary constraints on the AGB phase. In the first
paper of this series \citep[][hereafter referred to as "Paper
I"]{guandalini} we analyzed a sample of C stars reconstructing
their SEDs up to 45 $\mu$m on the basis of space-borne infrared
observations from the ISO and MSX missions. We found evidence for
a relatively high average C-star luminosity, suggesting that
the so-called "C-star luminosity problem" \citep{cohen} could be
simply an effect of poor estimates of the luminosity, due to
insufficient knowledge of the emission at mid-infrared
wavelengths. In the same paper the available mass loss rates and
their correlation with infrared colors were also presented. In the
second paper \citep[][hereafter referred as "Paper II"]{guabus} we
extended the analysis to the MS and S giants, where the
enhancement of Carbon is more moderate than in C stars. We
examined there the "luminosity problem" only; here we want to combine these
results with an analysis of stellar winds, thanks to the recent
analysis made by \citet{ramstedt09}.
This research will be completed in a forthcoming work where we will examine
the luminosities and mass loss rates of galactic M-type AGB stars
in relation to the available infrared observations.

In Sect. \ref{sect2} we present the sample of studied stars and
discuss the choices made in selecting and organizing the sources
according to the quality of the available data. In Sect.
\ref{sect3} we analyze mass loss rates as functions of various
parameters (infrared colors, periods of variability, bolometric
luminosity) looking for quantitative relations that will be
commented upon keeping in mind the results of Paper II. Finally,
in Sect. \ref{sect4} some preliminary conclusions are drawn.

\section{The sample \label{sect2}}

In the selection of the sample we started from the compilation of
galactic S-type AGB sources presented in Paper II. We chose to
use only sources for which reliable estimates of the distance are
available, therefore only sub-samples A, B and C from that work
were considered (the only exception is in Fig. \ref{fig3}).
We searched these sub-samples for sources with reliable mass loss rates obtained
through procedures similar to those suggested by \citet{knappmorris}. The
adopted estimates were taken from different sources in the literature and have been
all updated with new estimates of the distances. All these sources
have, as a basis, radio observations of the CO lines. Whenever
possible, we used the data from \citet{ramstedt,ramstedt09},
obtained with a procedure involving the radiative transfer model
presented in \citet{schoier,olof02,ramstedt08}. As a second option,
we selected the expressions from \citet{loup} applied also by
\citet{winters03}. In this case a system of two equations permits
us to evaluate both the mass loss rate and the CO
photo-dissociation radius (see the references given above for more
details). The results are listed in Table \ref{table:1}.

In the Table the stars are divided in three subgroups separated
with horizontal lines. In the first subgroup \textbf{we can see} all the sources
of the spectral type S-MS. The second is made of four SC stars. In the last subgroup we have a single
source with an uncertain classification, variously estimated as being
O-rich or C-rich. For this last star (\object{VX Aql}) the data in the table are
derived by considering it as being O-rich. The variability types
have been taken from the GCVS catalogue \citep{samus}, while the
distances are taken from Paper II. These estimates were derived
in two ways: a) from the revised Hipparcos parallaxes
\citep{vleu2007}; b) from the period-luminosity relations for
O-rich AGBs (see the appendix in Paper II for a detailed description).

\begin{figure*}[t!!]
\centering \resizebox{\textwidth}{!}{\includegraphics{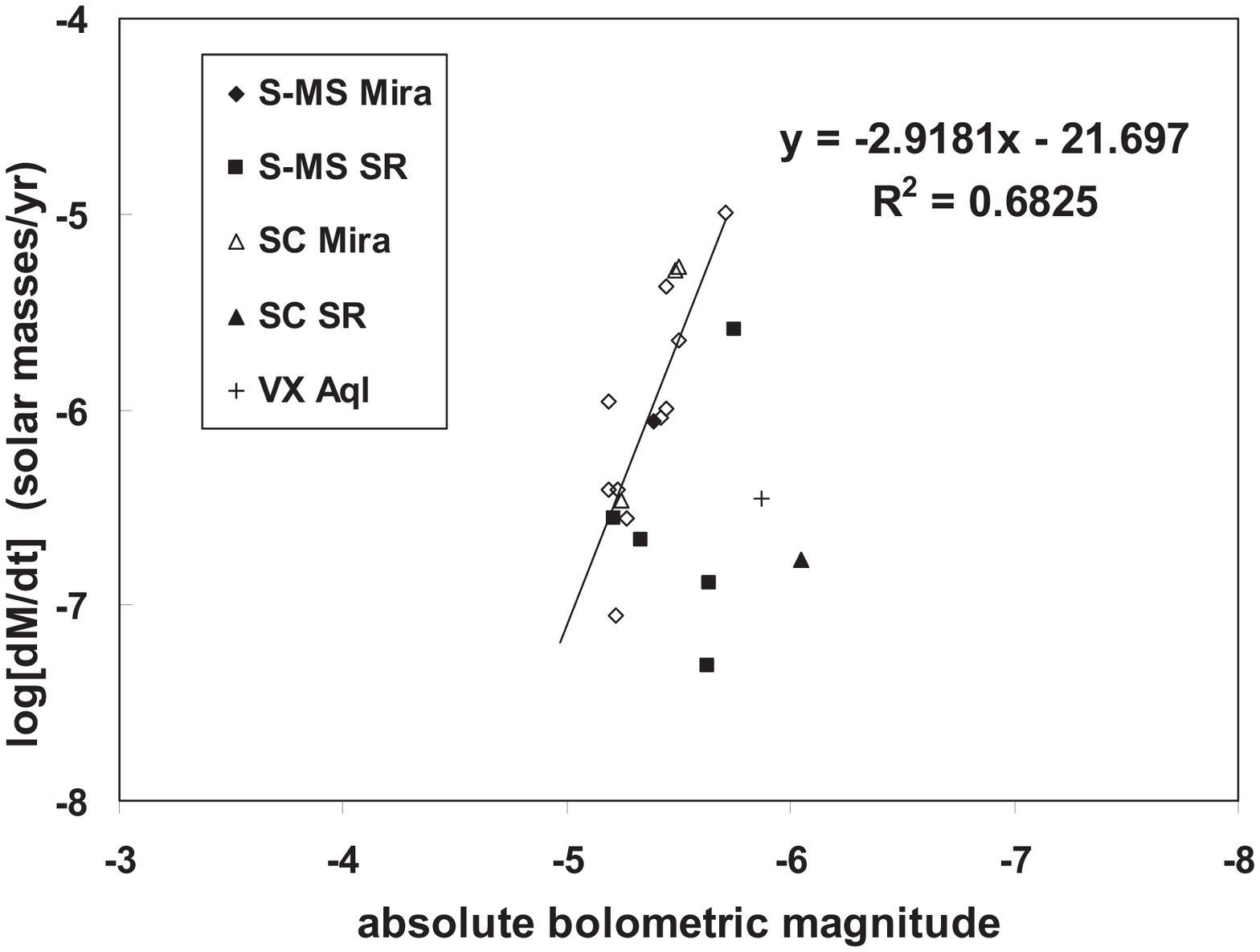}
\includegraphics{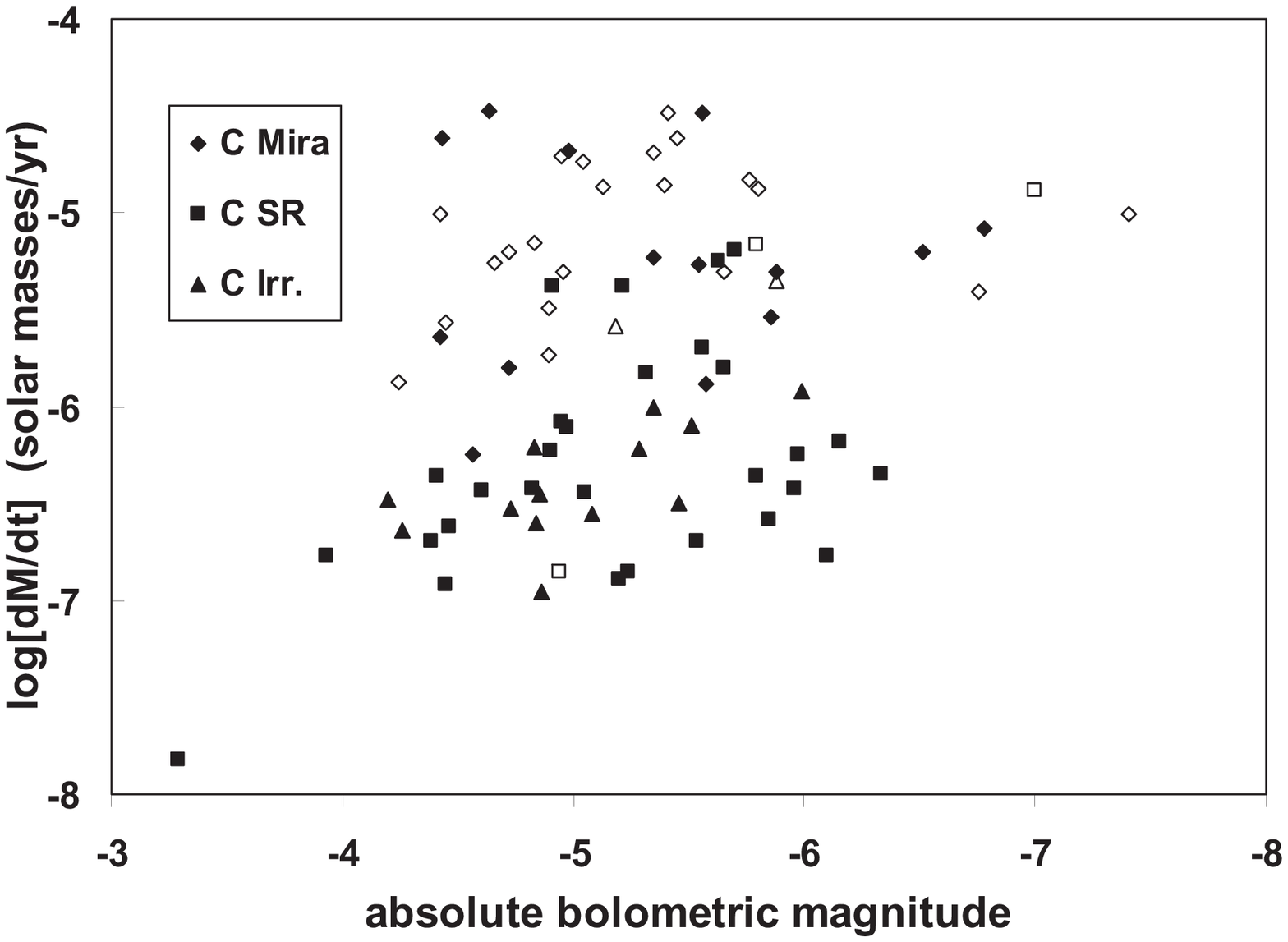}}
\caption{Left panel: the mass loss rates for S-type stars plotted as a
function of their absolute bolometric magnitudes. Here and in the
other plots of Fig. \ref{fig2}: 1) rhombs and squares are Mira
and Semiregular S and MS stars respectively; 2) full symbols indicate sources
with distance estimates derived from the Hipparcos parallaxes; empty ones are
Miras whose distance has been estimated with the period-luminosity relations;
3) empty triangles are SC Miras with the distances calculated through the
period-luminosity relations; the unique full triangle is a SC Semiregular with the
distance obtained with the Hipparcos parallax. The estimates of the bolometric
magnitudes come from Paper II. The least-square fit considers only S and MS Miras.
Right panel: the mass loss rates for C-rich stars plotted as a
function of absolute bolometric magnitudes (data from Paper I). Sources shown with a
full symbol are from Table 1 of Paper I, while sources indicated
with an empty symbol are from Table 2 of the same paper.} \label{fig1}
\end{figure*}

\begin{figure*}[t!!]
\centering \resizebox{\textwidth}{!} {\includegraphics{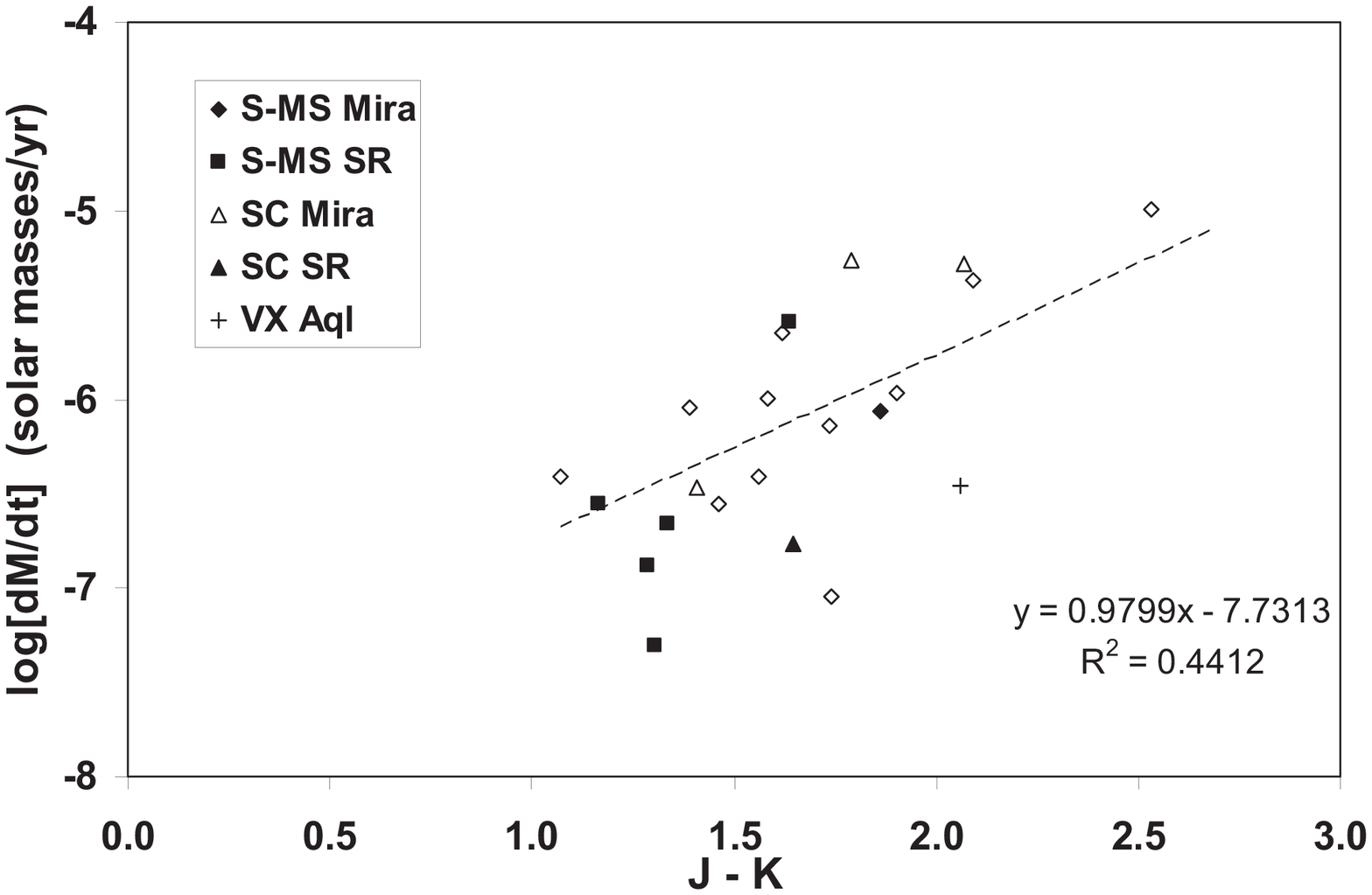}
\includegraphics{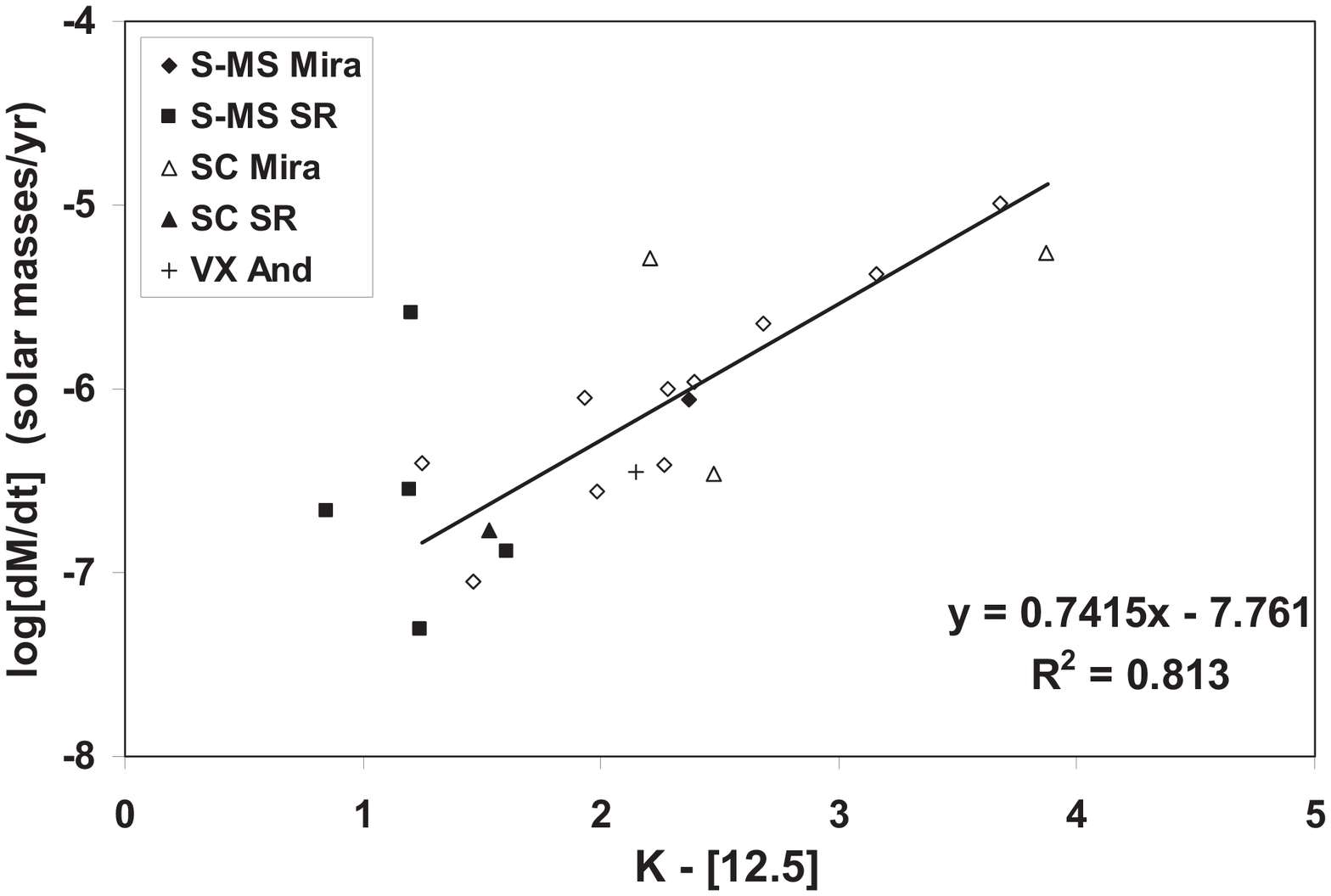}}
\caption{Relations linking the mass loss rates of S-type AGB
sources to the IR colors. The photometric data have been taken
from Paper II (where a detailed analysis can be found) and come
respectively from 2MASS for the J and K bands and from ISO-SWS,
MSX and IRAS-LRS for the [12.5] band. Here and elsewhere "K" is a
compact notation for the 2MASS filter K$_{s}$ (K-short). More
details about the two plots can be found in the text. } \label{fig2}
\end{figure*}

Finally, we give a suggestion for the nature of the stars in our sample
(either intrinsic or extrinsic). These indications come from Paper II,
\textbf{where we describe how they were obtained}.
[originally, they derive from \citet{vaneck00} and \citet{yang}].
Apart from a few uncertain cases, the sources considered in this case are all
intrinsic AGB stars.

\section{Mass loss rates and other physical parameters \label{sect3}}

We now compare the mass loss rates presented in Table
\ref{table:1} with other relevant physical parameters. We apply
the results published in Paper II and Paper I.

In Fig. \ref{fig1} (left panel) we examine mass loss rates and absolute
bolometric magnitudes. The luminosities were calculated in
Paper II with the bolometric corrections presented there
and obtained by studying the spectral energy distributions of
available S-type stars up to 45 $\mu$m. Mira sources of S-MS type
are located inside a relatively narrow region. For these stars we can see
that the mass loss rates increase when the luminosity increases. If we make a linear
least-square fit (see Fig. \ref{fig1}), we obtain a correlation coefficient of $R^{2} = 0.6825$. We
have to remind the reader, as a word of caution, that the sample used is rather
small, which may lead to some selection biases. However, if we
look at the whole sample of $\sim$600 sources from Paper II, we recall that those
examined here are a representative sub-sample for Mira S-MS stars.
Moreover, they are all part of the best studied sub-samples from Paper II.
The SC sources are located in the same region of the plot, but we decided to not study
them together with S and MS stars because of their peculiar chemical properties in
the circumstellar envelopes, like the presence of compounds containing \textbf{Carbon}.
Mass loss rates for S stars reach values of almost $10^{-5}$
M$_{\odot}$/yr. Although the highest values observed for C-rich
stars are not reached (see Fig. \ref{fig1}, right panel and Paper I),
nonetheless the rates we find are high enough, and also associated with high values of the
luminosity, to again suggest that the evolutionary scheme
MS$-$S$-$SC$-$C should not be taken as granted. The most luminous S-type
stars with the strongest stellar winds might be already close to the
AGB end, and would probably never become C-rich (see Paper II for
more details on this topic).

As we compare the two panels of Fig. \ref{fig1} we observe
that the range of luminosities for this sample of S-MS stars is very narrow,
around $-$5. Instead, C stars from Paper I span a much wider range of
bolometric luminosities, and, if considered together, they do not
show a clear correlation between \textbf{absolute} bolometric magnitudes and
mass loss rates. However, we recall that in Paper I a revision of the
parallaxes from the Hipparcos original release \citep{bergeat} was used
instead of the completely new release \citep{vleu2007} adopted here. Moreover,
the crucial work from \citet{white06} regarding the period-luminosity
relations for Galactic Carbon Miras was published a few months after
Paper I and its results could not be integrated in our previous work. The large scatter of C-rich
sources observed in the right panel could be caused by poorer estimates
of their distances. Finally, if we consider our entire sample of S-type stars
from Paper II, we do not find sources totally obscured by dust. Yet
some of them have a considerable IR excess: for instance, S Cas (observed by
ISO-SWS) has the peak of luminosity in its spectral energy distribution at
around 10 $\mu$m. The same existence of totally-obscured S stars is doubtful, as they are
on average of smaller mass than C stars (see Paper II) and might therefore lose mass at lower
rates \textbf{and cross} the final superwind stage only briefly.

The estimates of \textbf{absolute} bolometric magnitudes and mass loss rates are not always independent.
They are so only if the distance is derived from Hipparcos parallaxes, but unfortunately
a Hipparcos distance is available only for one Mira of our sample. Where
the distance is derived from the period-luminosity relations, the absolute luminosity and rates of
the stellar winds are necessarily correlated. Indeed, the distance is derived from the apparent
bolometric magnitude (obtained thanks to the bolometric corrections presented in Paper II)
and the period-luminosity relations. This distance is then used to re-scale
mass loss rates found in the literature.

We can try to disentangle the problem of the independence between these two parameters
by checking the behavior of the mass loss rates if compared with parameters that are
distance-independent: variability period and infrared colors.

1) \textbf{The period.} If we compare the mass loss rates with the
variability periods (taken from Paper II) we find again a correlation between
S-MS Miras and a coefficient of correlation for a linear least-square fit $R^{2} = 0.6777$.
This is obviously expected, because almost all the sources involved in the fit
have the distances derived from period-luminosity relations. The crucial thing that
should be pointed out here is that for S-MS Mira stars there are signs of a correlation
between mass loss rates and periods of variability. One could wonder if the same thing
happens also for \textbf{Semiregular} sources at shorter periods, but in our case the sample
is too small and scattered to give reliable indications.

2) \textbf{The color.} Near-to-mid infrared colors are crucial
to study the main parameters of AGB stars: for both
C-rich (Paper I) and S-type (Paper II) stars our best bolometric
corrections were obtained with these colors. Therefore, the
bolometric magnitudes depend on them.
We examine now Fig. \ref{fig2}. It shows two plots in which the mass
loss rates are compared with two different colors:
a near-IR color (J-K) in the left panel and a near-to-mid IR one
(K-[12.5]) in the right panel. This figure can give us important informations:
it confirms the argument made until now and offers us indications \textbf{about}
which wavelengths are more useful in the search for relations between the stellar winds and
the photometric colors (and the temperature).

\begin{figure*}[t!!]
\centering \resizebox{\textwidth}{!}
{\includegraphics[height=4.6cm,width=6cm]{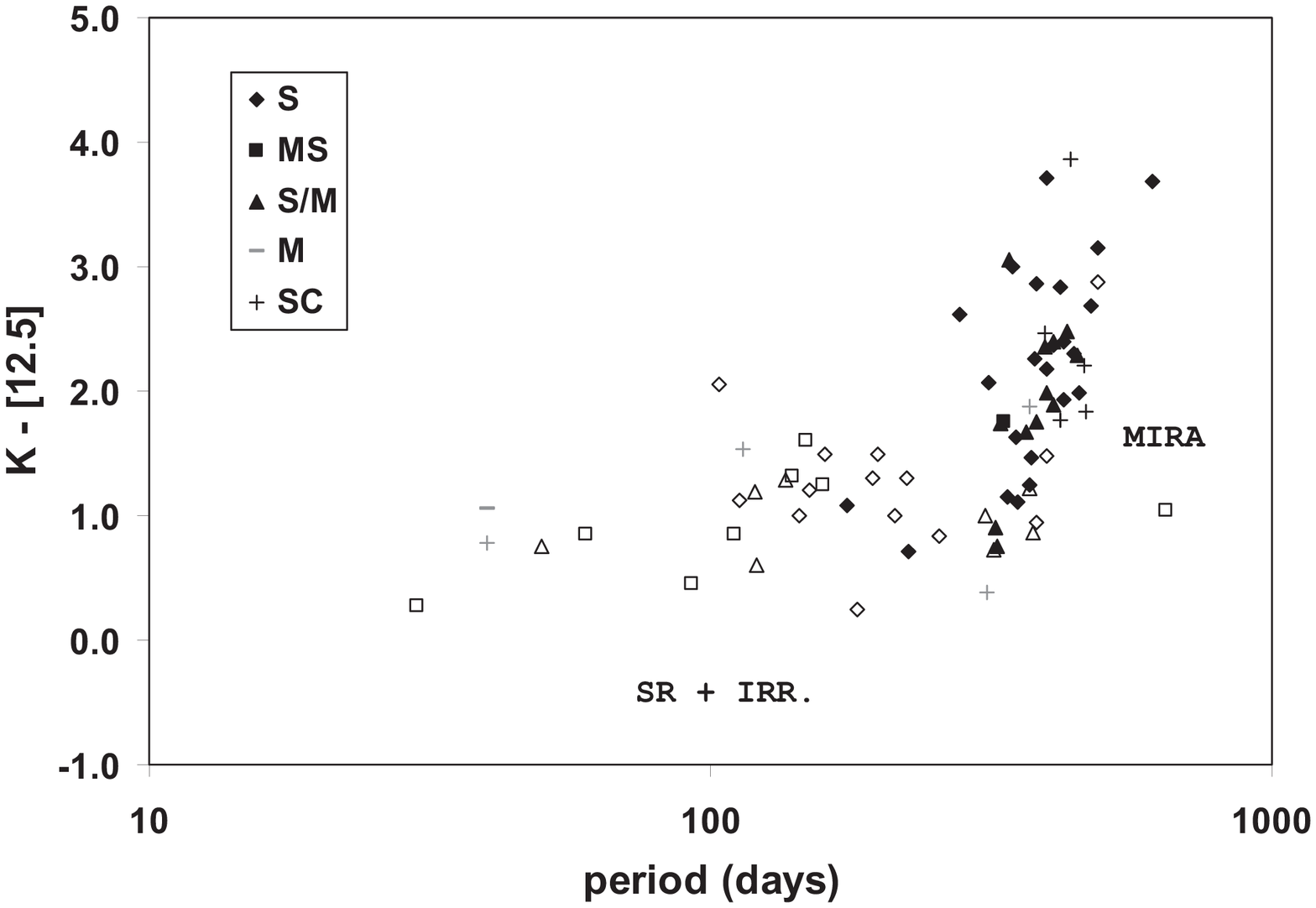}
\includegraphics[height=4.55cm,width=6cm]{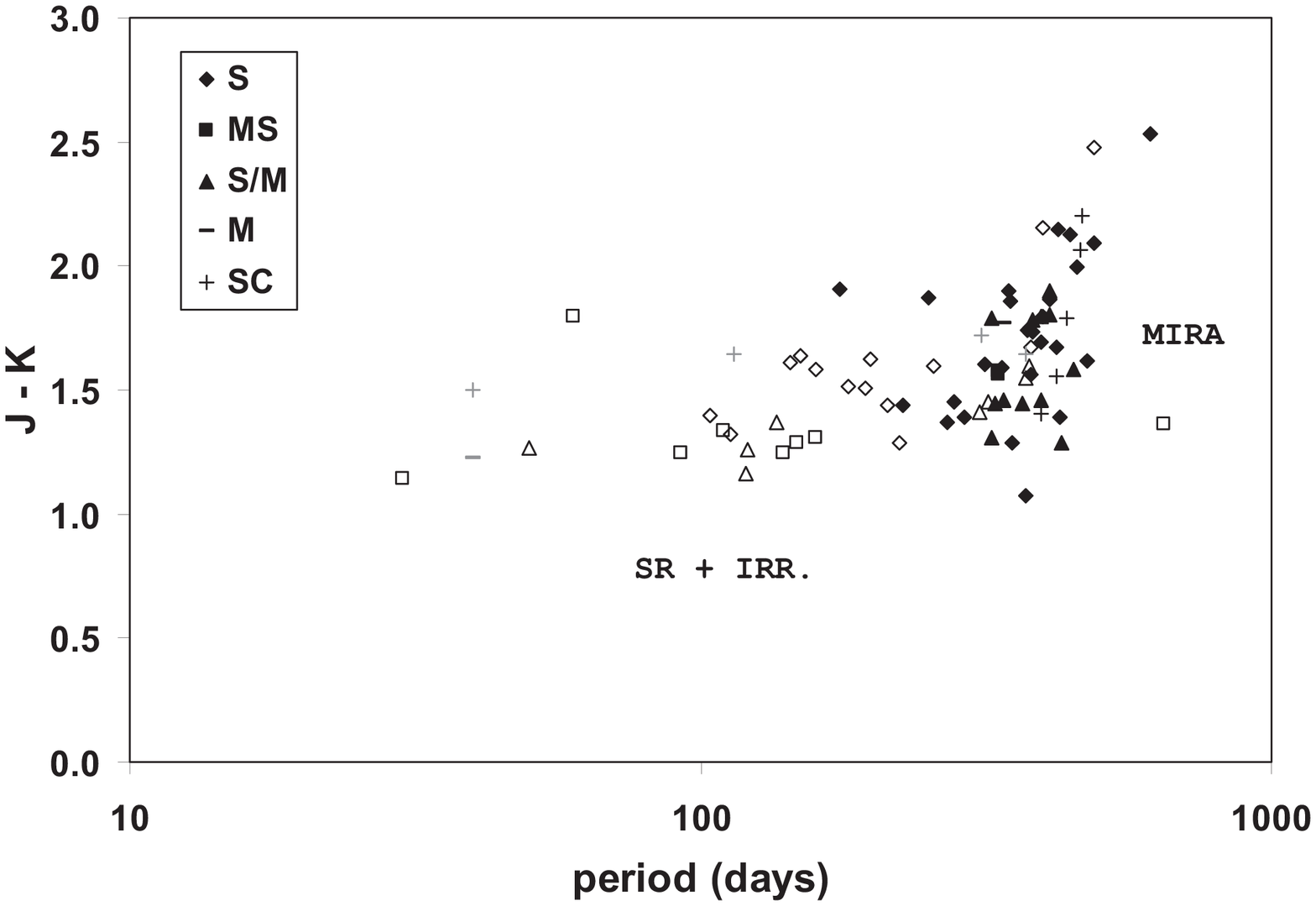}}
\caption{A near-to-mid infrared color (left panel) and a
near-infrared one (right panel) plotted as a function of period of
variability. In this case full dots are Miras, while empty dots are
Semiregulars and Irregulars. Sources are from the Paper II sample. In both panels the regions
of the Miras and of the Semiregulars/Irregulars are indicated. Photometric data have
been taken from Paper II, periods come from the GCVS catalogue \citep{samus}.} \label{fig3}
\end{figure*}

The K-[12.5] color adopted in the right panel of Fig.
\ref{fig2} was also one of the baselines used to derive bolometric corrections
for S stars in Paper II. Therefore this color is as important as the variability period
to infer absolute luminosities (and distances). S-MS Mira
stars show a well-defined relation between mass loss rates and
both mentioned parameters. In the right panel of Fig.
\ref{fig2} we have added a least-square fit based on
these sources. The results are indicated in the plot. The tendency
to have stronger mass loss rates for redder sources is clear, and the
linear relation has a good reliability \textbf{($R^{2} = 0.813$)}.
Mass loss rates for S-MS Miras show again a clear correlation with another parameter (the
near-to-mid IR color) calculated in an independent way.
We caution the reader again that Mira sources
show some variability also in the 10 $\mu$m region \citep{busso07},
and this could affect our results.

We can also examine the different results that we obtained in Fig.
\ref{fig2} if we use a near-IR color [(J-K) in the left panel]
or a near-to-mid infrared color [(K-[12.5]) in the right panel] as
a function of the mass loss rates. As already shown in Paper I, the
different stellar classes (Miras, Semiregulars, etc.) are
best separated when mid-IR wavelengths are
included in the colors. In the left panel (J-K color),
the Miras seem to show a relation, but the data are quite scattered (see the
coefficients of correlation in the plots);
instead, in the right panel (K-[12.5] color) the least-square line that we computed for
the Mira S-MS sources has a much better coefficient of
correlation than in the previous case. On the other hand,
Semiregulars are very scattered in both plots, without signs of clear
patterns in their behavior. In Paper I we pointed out that the
entity of the overlapping region between Miras and Semiregulars could
vary with the colors in abscissa. In this case Miras and Semiregulars are
again quite well separated in both plots, and this effect is best
enhanced if a near-to-mid infrared color is adopted.

The pattern between mass loss rates and luminosities shown in
Fig. \ref{fig1} originates from the relations with two independent parameters:
periods of variability and near-to-mid infrared colors (Fig. \ref{fig2}, right panel).
This pattern should not be contaminated by errors in the selection
criteria. Moreover, we can examine the plots in Fig. \ref{fig3} where these two parameters
(the colors J-K and K-[12.5] and the periods of variability) are compared.
In this Figure, as these parameters are all independent from the distance of the sources,
we consider all the sources from Paper II \textbf{that have} both an estimate of the period and
reliable mid-IR observations (see Paper II). From the plots we see that, if associated
with a well-chosen color, the period of variability separates the
Miras population in the best way from the Semiregulars. Moreover (left panel), when we use a near-to-mid infrared
color the Mira subgroup seems to show a correlation between K-[12.5] and the period of variability.
Unfortunately, in this case the data for Miras are more scattered
(the coefficient of correlation of a linear least-square fit for full rhombs and squares is $R^{2} = 0.346$),
nonetheless, there is the hint of a pattern, because sources with longer periods are on average
redder. We have to remember that the classification of variables between Miras and Semiregulars does not
have a physical basis: there are lot of sources in a "grey zone" that could be (and have been) classified
in both ways depending on the light-curves used. Looking at the right panel of Fig. \ref{fig3},
we see that again a plot with pure near-IR colors has the data more scattered than a
similar plot with near-to-mid infrared colors. This underlines the importance of considering
both the cool, dusty, circustellar envelope (mid-IR) and the
warmer regions near the central star (near-IR) in a photometric
analysis for these sources. Colors derived only from
near-infrared fluxes are not sensitive enough (and do not have a
long enough baseline) to give a good description of the relations
between colors and other crucial physical parameters like periods
and mass loss rates.

Until now we have considered only the mass loss rates from
the gaseous part of the circumstellar envelopes. Completing this
analysis requires a similar study on the mass loss rates for the
dust component and an estimate of the dust-to-gas ratios. We have checked
the dust mass loss rates from \citet{ramstedt09} and \citet{groenewegen} and have
calculated the dust-to-gas ratios from them and from our estimates of gas mass loss
rates. The ratios are reported in the last column of Table \ref{table:1}.
They are on average lower than the ones observed
in the interstellar medium. This raises doubts on their reliability (and particularly
on the dust mass loss rates), because AGB envelopes should be among
the sites where dust is created. Further researches should be performed
on this topic.

\begin{figure}[t!!]
\centering \resizebox{\hsize}{!} {\includegraphics{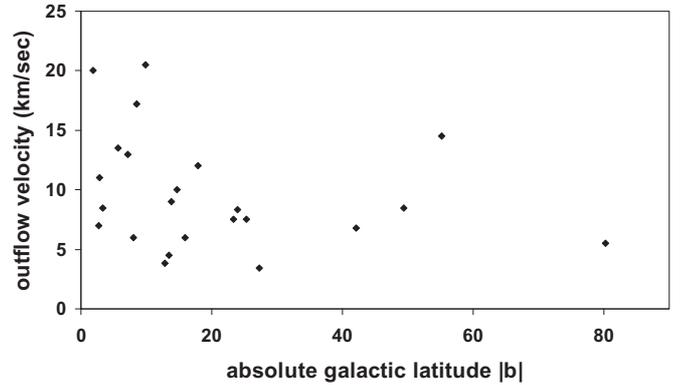}}
\caption{Plot of the circumstellar outflow velocity v$_{e}$ compared with the
absolute value of the galactic latitude $|b|$.} \label{fig4}
\end{figure}

Outflow velocities have also been included in Table \ref{table:1}.
They can be measured from the CO profiles used to estimate the mass loss
rates for gas. \citet{zucke89} in their Fig. 3 observed a correlation
between outflow velocities and absolute values of the
galactic latitude $|b|$ for a sample of C-rich sources: the ones with an
higher outflow velocity ($>$ 20 km/s) have a low value of $|b|$ ($<$ 20 degrees),
while the ones with an high value of $|b|$ have a low outflow velocity.
This property, if confirmed, could be linked to the age and mass
of the stars, giving us interesting hints on their evolutionary properties.
In Fig. \ref{fig4} we present a similar plot made with our sample of S-type
stars. The same relation between outflow velocity and galactic latitude $|b|$ appears
also in this case. Moreover, the data in Fig. \ref{fig4} suggest that the absence
of really high-mass-loss stars in our sample seems to be
linked with the absence of sources with a high outflow velocity.
This finding, if linked with their quite high values of luminosity and mass loss rate
(even if the highest values of the C-rich stars of Paper I are not reached), confirms
that S-type stars in our Galaxy have masses slightly lower than those of \textbf{Carbon} sources.
A good fraction of them could evolve along the AGB phase without becoming C-rich
(see Paper II for more details).

\section{Conclusions \label{sect4}}

We integrated the analysis of MS-S-SC galactic AGB
stars, started in Paper II, by examining the mass loss rates for a
sample of sources with reliable estimates of the distance.
Selection criteria for the estimates of the stellar winds were
discussed and the chosen rates were updated according to
new measurements of distances.

The figures shown in our analysis suggest the existence
of a correlation between mass loss rates and absolute
bolometric magnitudes for Mira variables (Fig. \ref{fig1}). In the past the
existence of such a correlation was doubtful \citep[see e.g.][and
references therein]{vanloon99a,vanloon99b,vanloon2005} but an
accurate selection of the sample appears to provide enough evidence (for Mira-type sources).
It appears to be linked with and to \textbf{be} originated from
relations that the rates of the stellar winds have
with two physical parameters that are independent from the distance:
the \textbf{period of variability} and the \textbf{K-[12.5] color}
(Fig. \ref{fig2}, right panel).

These two parameters are fundamental tools for the two methods
adopted to estimate the absolute bolometric luminosities in
Paper II: the period-luminosity relations and the bolometric
corrections.

Figure \ref{fig2} shows a strong linear relation between
the mass loss rates and the K-[12.5] color for Mira sources.
This relation could be of the utmost importance,
because, through its application, we could obtain reliable
estimates of the mass loss rates directly from a photometric
color. In this way the rates of the stellar winds for
Mira sources could be estimated without the need of radio observations of the CO
lines (or other independent observations) and/or estimates
of the distance. Near-infrared colors (i.e. J-K) seem to
be less useful than the near-to-mid infrared ones.

Finally, if we directly compare the period of variability with
the near-to-mid infrared color K-[12.5], we find that this is the best color index
to separate Miras and Semiregulars. Moreover, a relation might exist between
these two parameters for Miras, but the scatter in our sample is too large to draw
final conclusions on it.

\begin{acknowledgements}
This research was supported by the Italian Ministry of Research
under contract PRIN2006-022731. The author thanks the referee,
Dr. C. Loup, for an extensive and helpful review. The author also
thanks M. Busso for many clarifying discussions.
\end{acknowledgements}


\begin{thebibliography}{}

\bibitem[Bergeat \& Chevallier(2005)]{bergeat}
    Bergeat, J., \& Chevallier, L. 2005, A\&A, 429, 235

\bibitem[Busso et al.(1999)]{busso99}
    Busso, M., Gallino, R., \& Wasserburg, G.J. 1999, ARA\&A, 37, 239

\bibitem[Busso et al.(2007)]{busso07}
    Busso, M., Guandalini, R., Persi, P., Corcione, L., \& Ferrari-Toniolo, M.
    2007, AJ, 133, 2310

\bibitem[Cohen et al.(1981)]{cohen}
    Cohen, J.G., Persson, S.E., Elias, J.H., \& Frogel, J.A. 1981,
    ApJ, 249, 481

\bibitem[Groenewegen \& de Jong(1998)]{groenewegen}
    Groenewegen, M.A.T., \& de Jong, T. 1998, A\&A, 337, 797

\bibitem[Guandalini \& Busso(2008)]{guabus}
    Guandalini, R., \& Busso, M. 2008, A\&A, 488, 675  (Paper II)

\bibitem[Guandalini et al.(2006)]{guandalini}
    Guandalini, R., Busso, M., Ciprini, S., Silvestro, G., \& Persi, P. 2006, A\&A, 445,
    1069  (Paper I)

\bibitem[Herwig(2005)]{herwig}
    Herwig, F. 2005, ARA\&A, 43, 435

\bibitem[Knapp \& Morris(1985)]{knappmorris}
    Knapp, G.R., \& Morris, M. 1985, ApJ, 292, 640

\bibitem[Loup et al.(1993)]{loup}
    Loup, C., Forveille, T., Omont, A., \& Paul, J.F. 1993, A\&AS,
    99, 291

\bibitem[Olofsson et al.(2002)]{olof02}
    Olofsson, H., Gonz\'{a}lez Delgado, D., Kerschbaum, F., \& Sch\"{o}ier, F.L.
    2002, A\&A, 391, 1053

\bibitem[Ramstedt et al.(2006)]{ramstedt}
    Ramstedt, S., Sch\"{o}ier, F.L., Olofsson, H., \& Lundgren,
    A.A. 2006, A\&A, 454, L103

\bibitem[Ramstedt et al.(2008)]{ramstedt08}
    Ramstedt, S., Sch\"{o}ier, F.L., Olofsson, H., \& Lundgren,
    A.A. 2008, A\&A, 487, 645

\bibitem[Ramstedt et al.(2009)]{ramstedt09}
    Ramstedt, S., Sch\"{o}ier, F.L., \& Olofsson, H. 2009, A\&A, 499,
    515

\bibitem[Salpeter(1974)]{salpeter}
    Salpeter, E.E. 1974, ApJ, 193, 585

\bibitem[Samus et al.(2004)]{samus}
    Samus, N.N., Durlevich, O.V., et al. 2004, VizieR On-line Data Catalog: II/250,
    Institute of Astronomy of Russian Academy of Science and
    Sternberg, State Astronomical Institute of the Moscow State University, 2004

\bibitem[Sch\"{o}ier \& Olofsson(2001)]{schoier}
    Sch\"{o}ier, F.L., \& Olofsson, H. 2001, A\&A, 368, 969

\bibitem[Sedlmayr(1994)]{sedlmayr94}
    Sedlmayr, E. 1994, Lecture Notes in Physics, 428, 163

\bibitem[Straniero et al.(2003)]{straniero}
    Straniero, O., Dom\'{i}nguez, I., Cristallo, S., \& Gallino,
    R. 2003, PASA, 20, 389

\bibitem[Uttenthaler et al.(2007)]{utt07}
    Uttenthaler, S., Hron, J., Lebzelter, T., et al. 2007, A\&A, 463, 251

\bibitem[Van Eck et al.(2000)]{vaneck00}
    Van Eck, S., Jorissen, A., Udry, S., et al. 2000, A\&AS, 145,
    51

\bibitem[van Leeuwen(2007)]{vleu2007}
    van Leeuwen, F. 2007, A\&A, 474, 653

\bibitem[van Loon et al.(1999a)]{vanloon99a}
    van Loon, J.Th., Zijlstra, A.A., \& Groenewegen, M.A.T. 1999a, A\&A, 346, 805

\bibitem[van Loon et al.(1999b)]{vanloon99b}
    van Loon, J.Th., Groenewegen, M.A.T., de Koter, A., et al. 1999b, A\&A, 351, 559

\bibitem[van Loon et al.(2005)]{vanloon2005}
    van Loon, J.Th., Cioni, M.-R.L., Zijlstra, A.A., \& Loup, C. 2005, A\&A, 438, 273

\bibitem[Wachter et al.(2002)]{wachter02}
    Wachter, A., Schr\"{o}der, K.-P., Winters, J.M., Arndt, T.U., Sedlmayr, E. 2002, A\&A, 384, 452

\bibitem[Whitelock et al.(2006)]{white06}
    Whitelock, P.A., Feast, M.W., Marang, F., \& Groenewegen,
    M.A.T. 2006, MNRAS, 369, 751

\bibitem[Winters et al.(2003)]{winters03}
    Winters, J.M., Le Bertre, T., Jeong, K.S., Nyman, L.-\AA., \& Epchtein, N.
    2003, A\&A, 409, 715

\bibitem[Yang et al.(2006)]{yang}
    Yang, X., Chen, P., Wang, J., \& He, J. 2006, AJ, 132, 1468

\bibitem[Zuckerman \& Dick(1989)]{zucke89}
    Zuckerman, B., \& Dyck, H.M. 1989, A\&A, 209, 119

\end{thebibliography}
\end{document}